\documentclass[12pt]{article}
\usepackage{epsfig,a4}

\newcommand{\mgut}{\ensuremath{\mathrm{M_{GUT}}}}

\newcommand{\Mone}{\ensuremath{\mathrm{M_1}}}
\newcommand{\Mtwo}{\ensuremath{\mathrm{M_2}}}

\def\ga{\mathrel{\raise.3ex\hbox{$>$\kern-.75em\lower1ex\hbox{$\sim$}}}}
\def\la{\mathrel{\raise.3ex\hbox{$<$\kern-.75em\lower1ex\hbox{$\sim$}}}}
\newcommand{\gev}{\ensuremath{\mathrm{GeV}}}

\newcommand{\Tcsq}{\ensuremath{\mathrm{TeV}/c^2}}
\newcommand{\Gcsq}{\ensuremath{\mathrm{GeV}/c^2}}

\newcommand{\rts}{\ensuremath{\sqrt{s}}}

\newcommand{\ocdm} {\Omega_\mathrm{CDM}}

\newcommand{\bkin}{\ensuremath{\mathrm{\beta_{kin}}}}
\newcommand{\bkincube}{\ensuremath{\mathrm{\beta^3_{kin}}}}
\newcommand{\Pd}{\ensuremath{\mathrm{d}}}
\newcommand{\Pu}{\ensuremath{\mathrm{u}}}

\newcommand{\Pad}{\ensuremath{\overline{\mathrm{d}}}}
\newcommand{\Pau}{\ensuremath{\overline{\mathrm{u}}}}

\newcommand{\epem}{\ensuremath{\mathrm{e^+e^-}}}

\newcommand{\ggqq}{\ensuremath{\gamma\gamma \rightarrow \mathrm{q}\overline{\mathrm{q}}}}
\newcommand{\ggff}{\ensuremath{\gamma\gamma \rightarrow \mathrm{f}\overline{\mathrm{f}}}}

\newcommand{\ffbar}{\ensuremath{\mathrm{f}\overline{\mathrm{f}}}}
\newcommand{\ffff}{\ensuremath{\epem\!\to\!\mathrm{f_1f_2}\overline{\mathrm{f_1}}\overline{\mathrm{f_2}}}}

\newcommand{\toto}{\ensuremath{\tau^{+}\tau^{-}}}
\newcommand{\zz}   {\ensuremath{\mathrm Z\gamma^{*}}}

\newcommand{\wen}{\ensuremath{\mathrm{We\nu}}}
\newcommand{\znn}{\ensuremath{\mathrm{Z\nu\nu}}}
\newcommand{\ww}{\ensuremath{\mathrm{W^+W^-}}}

\newcommand{\gagall}{\ensuremath{\gamma\gamma\rightarrow \ell\ell }}

\newcommand{\tanb}{\ensuremath{\tan\beta}}

\newcommand{\mzero}{\ensuremath{m_{0}}}

\newcommand{\Pcha}{\ensuremath{\chi^{\pm}}}

\newcommand{\Pccp}{\ensuremath{\chi^+}}
\newcommand{\Pccm}{\ensuremath{\chi^-}}

\newcommand{\PChiz}[1]{\ensuremath{\chi^0_{#1}}}
\newcommand{\lsp}{\ensuremath{\chi^0_{1}}}
\newcommand{\PSnu}{\ensuremath{\widetilde{\nu}}}

\newcommand{\PSl}{\ensuremath{\widetilde{\ell}}}

\newcommand{\PSe}{\ensuremath{\widetilde{\mathrm{e}}}}
\newcommand{\PSf}{\ensuremath{\widetilde{\mathrm{f}}}}
\newcommand{\PSmu}{\ensuremath{\widetilde{\mu}}}

\newcommand{\PSt}{\ensuremath{\mathrm{\widetilde{t}}}}

\newcommand{\Mcha}{\ensuremath{m_{\chi^\pm}}}
\newcommand{\Mchi}{\ensuremath{m_{\PChiz{1}}}}
\newcommand{\Mlsp}{\ensuremath{m_{\PChiz{1}}}}
\newcommand{\Msnu}{\ensuremath{m_{\tilde{\nu}}}}

\newcommand{\Msl}{\ensuremath{m_{\PSl}}}
\newcommand{\mA}{\ensuremath{m_{\mathrm A}}}
\newcommand{\mh}{\ensuremath{m_{\mathrm h}}}

\newcommand{\mZ}{\ensuremath{m_{\mathrm Z}}}

\newcommand{\mst}{\ensuremath{m_{\tilde{\mathrm t}}}}

\newcommand{\dm}{\ensuremath{\Delta M}}
\newcommand{\evis}{\ensuremath{E_{\mathrm{vis}}}}
\newcommand{\mvis}{\ensuremath{M_{\mathrm{vis}}}}

\newcommand{\notE}{\ensuremath{\mathrm{\not{\!\!E}}}}
\newcommand{\PT}{\ensuremath{P_T}}

\newcommand{\be}{\begin{equation}}
\newcommand{\ee}{\end{equation}}

 1
\font\elevenrm=cmr10 scaled\magstep 1
 1

\def\reff{\noindent}

\begin{document}

\begin{picture}(160,1)
\put(320,-10){\parbox[t]{45mm}{{\tt MPI-PhE/2000-24}}}
\put(320,-23){\parbox[t]{45mm}{{\tt hep-ex/0102013}}}
\end{picture}

\vspace*{1.8cm}
  \centerline{\bf DARK MATTER \& SUSY: LEP RESULTS 
  \footnote{
{\it Presented at Vulcano Workshop 2000: 
Frontier Objects In Astrophysics And Particle Physics, 
22-27 May 2000, Vulcano, Italy}}}
\vspace{1cm}
  \centerline{GERARDO GANIS}
\vspace{1.4cm}
  \centerline{MPI--PHYSIK 
  \footnote{\it Present address: INFN-Roma~2, Dipartimento di Fisica, 
             Universit\`a Roma ``Tor Vergata'', 
             Via Della Ricerca Scientifica 1, 00133 ROMA, ITALY}}
  \centerline{\elevenrm F\"ohringer Ring 6, 80805 M\"unchen, GERMANY}
  \centerline{E-mail: Gerardo.Ganis@cern.ch}
\vspace{2cm}
\begin{abstract}
The negative outcome of searches for supersymmetry performed at LEP
have been used to derive indirect constraints on the parameters of 
the most plausible supersymmetric candidates for cold dark matter, 
in particular for the lightest neutralino.
We review the basic ideas leading to the present lower limit on the lightest 
neutralino mass of  about 37 $\Gcsq$, 
with emphasis on the underlying assumptions. 
\end{abstract}

\section{ Introduction }

There are several hypotheses about the origin of the dark matter 
(see, for instance, Turner 1999). 
Among these, the possibility that the cold dark matter (CDM) is made, 
at least partially, of relic densities of elementary particles 
still survives if these particles are weakly interacting and massive 
(${\cal O}$(GeV) or heavier) (Ellis 1998, Bottino 1999). 
The enlarged particle spectrum found in extensions of the Standard Model (SM) 
generally accommodates for Weakly Interacting Massive Particles (WIMP)
that could become CDM candidates (Ellis 1998).
A large number of dedicated experiments have been designed to 
detect the signal expected on earth detectors if the Galaxy halo 
is made of such particles.  
Most of these experiments look for the signal generated by the 
recoiling nucleus after a WIMP has interacted in low noise detectors, 
usually located underground. 
 Other experiments look for WIMP annihilation leading to high energy 
neutrinos, high energy photons or antimatter in cosmic rays; 
for a review, see Feng 2000. 

A common feature of the experiments aiming at direct WIMP detection 
is the loss of sensitivity at low WIMP masses. For example, 
the first generation of Germanium experiments (Caldwell 1988) 
were sensitive to neutrino masses in the range 14 $\Gcsq$ to a few $\Tcsq$. 
On the other hand, a relatively light WIMP could be produced  
directly or via decay cascades in collisions at accelerators. 
In standard particle detectors such particles  would escape detection, 
just as neutrinos. The expected phenomenological signature would therefore 
be {\it missing energy}, i.e. effective energy-momentum non-conservation. 
The sensitivity of this indirect search for evidence of WIMP production is 
determined by three factors: 
{\it i)} the centre-of-mass energy ($\rts$), which sets the mass 
scale of the particle that can be produced in the collisions; 
{\it ii)} the effective WIMP production coupling, which 
determines the production cross sections and therefore the rate 
of events; this is typically larger than the nucleon-WIMP coupling 
relevant for the direct searches, because of the larger energy scale involved;
{\it iii)} the topology of the event, which determines the efficiency 
for detecting the interesting events; to be detected, in fact, WIMPs 
have to be produced in association with some visible, {\it triggering}, particle 
(Sect.~4). 
Once the triggering process is found, the indirect search has 
tipically good sensitivity in all the kinematically allowed range, 
therefore representing a potentially important complement to direct searches 
at small WIMP masses. 

The present accelerator experiments were not specifically optimized for  
this kind of searches, but have nonetheless some sensitivity 
to the predicted final states, mainly because of the good hermeticity which is 
the key point in determining the missing energy.  
Though the largest kinematical ranges are 
provided by hadronic colliders, the best sensitivity at present is  
reached by experiments at electron-positron colliders, which feature 
clean environments, allowing the study of all possible channels 
and their combined interpretation.  
In this paper we will focus on the CDM candidates predicted in 
supersymmetric extensions of the SM (reminded in Sect.~2) and 
on the constraints derived 
from the measurements done at the CERN LEP $\epem$ collider. 

\section{ Supersymmetry and its CDM candidates}

Supersymmetry (SUSY) is a symmetry relating bosons and fermions 
introduced in the building extensions 
of the SM to cure the so-called {\it hierarchy problem} 
(for reviews see Nilles 1984, Haber 1985, Martin 1997). 
The minimal particle content of a supersymmetric version of the 
standard model is shown in Table~1. Scalar partners ({\it sfermions}) are 
associated with ordinary matter fermions; the partners of gauge and 
Higgs bosons ({\it gauginos},{\it higgsinos}) are spin 1/2 fermions. 
In realistic supersymmetric models, the exact mass degeneracy between standard 
particles and supersymmetric partners is broken by the 
so-called {\it soft} supersymmetry breaking (soft-SB) terms, which preserve 
the theoretical benefits of these theories.  
The minimal set of parameters needed to specify the model contains two parameters 
of purely supersymmetric nature -  
the ratio between the vacuum expectation values 
(vev) for the two Higgs doublets, usually denoted as $\tanb$, and a mass parameter $\mu$ 
governing the mixing between the two Higgs doublets -  
and a few soft-SB ones:
three gaugino masses, $M_{i,i=1,2,3}$, associated with the 
$\mathrm{U(1)_Y,\ SU(2)_L}$ and $\mathrm{SU(3)_C}$ groups, respectively, to 
complete the description of the gaugino sector, 
one mass term $M_{\PSf}$ per scalar state  and 
the third family trilinear couplings $A_t,\ A_b,\ A_\tau$ for the sfermion sector.

\begin{table}[thb]
\begin{center}
\renewcommand{\arraystretch}{1.2}
{\small 
\begin{tabular}{|c|c||c|c|cc|cc|c|c|} 
\hline
Sector  &        & Quark & Lepton & \multicolumn{4}{|c|}{EW Gauge/Higgs bosons} & 
                   QCD & Gravity \\ 
\hline\hline
 SM      & Spin   & 1/2 & 1/2 & 1 & 0 & 1 & 0 & 1 & 2 \\ 
        &        & $(\!\!\begin{array}{c} u_L \\ d_L \end{array}\!\!),u_R,d_R$ 
                    & $(\!\!\begin{array}{c} \nu \\ l^-_L \end{array}\!\!),l^-_R$ 
                    & $\gamma,Z$ & $h,H,A$ 
                    & $W^\pm$ & $H^\pm$ & g & $G$ \\
\hline\hline
SUSY     & Spin   &  0  & 0 & \multicolumn{2}{|c|}{1/2} & \multicolumn{2}{|c|}{1/2} 
                    & 1/2 & 3/2 \\
         & Name   & Squark & Slepton & \multicolumn{2}{|c|}{Neutralino} &
                    \multicolumn{2}{|c|}{Chargino} & Gluino & \small Gravitino \\  
         &        & $(\!\!\begin{array}{c} \tilde{u}_L \\ \tilde{d}_L \end{array}\!\!),
                    \tilde{u}_R,\tilde{d}_R$ &
              $(\!\!\begin{array}{c} \PSnu \\ \tilde{l}^-_L \end{array}\!\!), 
                     \tilde{l}^-_R$ &
  \multicolumn{2}{|c|}{$\PChiz{i,i=1,...,4}$} & 
  \multicolumn{2}{|c|}{$\chi^\pm,\chi^\pm_2$} & 
        $\tilde{g}$  & $\tilde{G}$  \\
\hline
\end{tabular}
}
  \caption{{\small \em {Particle content of a Minimal Supersymmetric extension 
  of the SM}}}
 \end{center} 
\renewcommand{\arraystretch}{1.}
\end{table}

The large number of additional parameters is reduced by imposing the unification 
conditions at scale $\mgut$. For gauginos, at lowest order these read 
\be
\mathrm{M_1:M_2:M_3:M_{1/2}=\alpha_1:\alpha_2:\alpha_3:\alpha_{GUT}}
\ee
where $\mathrm{m_{1/2}}$ and $\mathrm{\alpha_{GUT}}$ are a common gaugino mass 
term and the unifying coupling constant at $\mgut$.
For sfermions, the unification relations read 
\be
\mathrm{M^2_{\PSf_i}=m^2_0+C_i m^2_{1/2}+D_i(\tanb,M_Z,\theta_W)}
\ee
where $\mzero$ is a common sfermion mass term at $\mgut$, and the C$_i$,D$_i$ 
quantities depend on the quantum numbers of the sfermion. The trilinear couplings 
are also derived from a common trilinear coupling $\mathrm{A_0}$ at $\mgut$.

The experimental non-observation of any barion number (B) or lepton number (L)
violation suggests that there is no tree-level violation of those quantities. 
This is enforced by requiring the conservation of the quantum number 
$\mathrm{R = (-1)^{3(B-L)+2S}}$
(S being the spin),  implying  that 
supersymmetric partners are produced in pairs, {\it i.e.} that the Lightest 
Supersymmetric Particle (LSP) is stable. 
Under the conservation of R parity, a neutral LSP is a natural candidate for 
CDM (Ellis 1983). By looking at Table 1 one can single out three possibilities:
{\it i) \underline{gravitinos}}: they are generally expected to be quite heavy, except in a special 
       class of models called Gauge-Mediated-SB models, where they are predicted 
       to be extremely light ($\sim$eV/c$^2$); in the latter case they would 
       contribute to hot DM, which is disfavored; we will not consider them 
       any longer;
{\it ii) \underline{sneutrinos}}: they can the LSP with the only free parameter the 
       mass $\mathrm{M_{\PSnu}}$; the interesting ranges for CDM 
       are $\mathrm{M_{\PSnu}}\sim$few GeV/c$^2$ or 
           $\mathrm{M_{\PSnu}}$ in [550,2300] GeV/c$^2$ (Ellis 1998); the latter 
       range is excluded by direct searches (Caldwell 1988);
{\it iii) \underline{neutralinos}}: the $\PChiz{1}$ is a flexible candidate 
       for CDM, giving the correct relic density $\ocdm h^2$ 
       in large portions of the parameter space
       (see, for instance, Ellis 2000a and references therein), as 
       can be seen in Fig.~1a. 

\begin{figure}[thb]
 \begin{center}
  \begin{tabular}{cc}
  \mbox{\epsfysize=8cm \epsffile{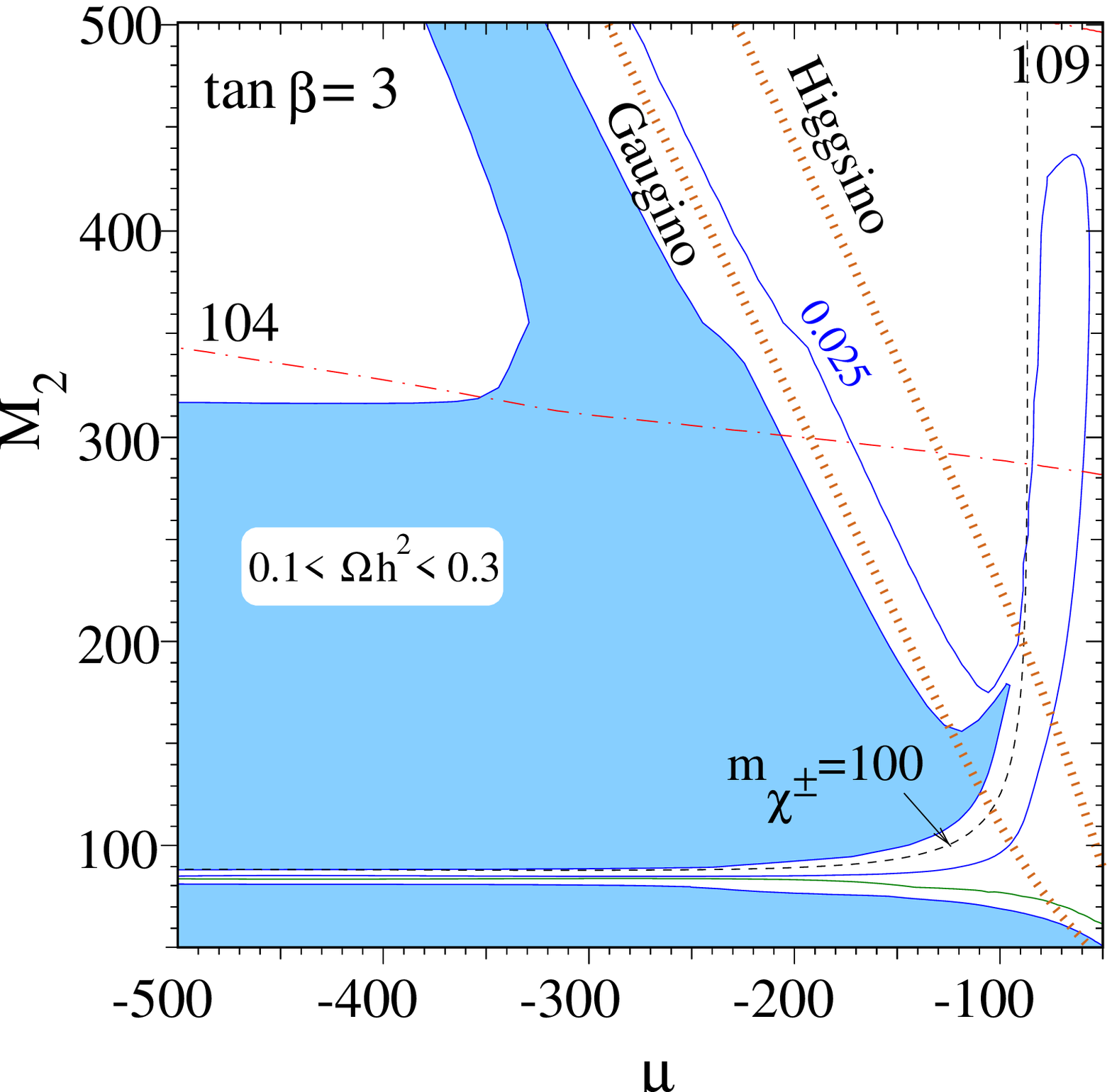}} &
  \mbox{\epsfysize=8cm \epsffile{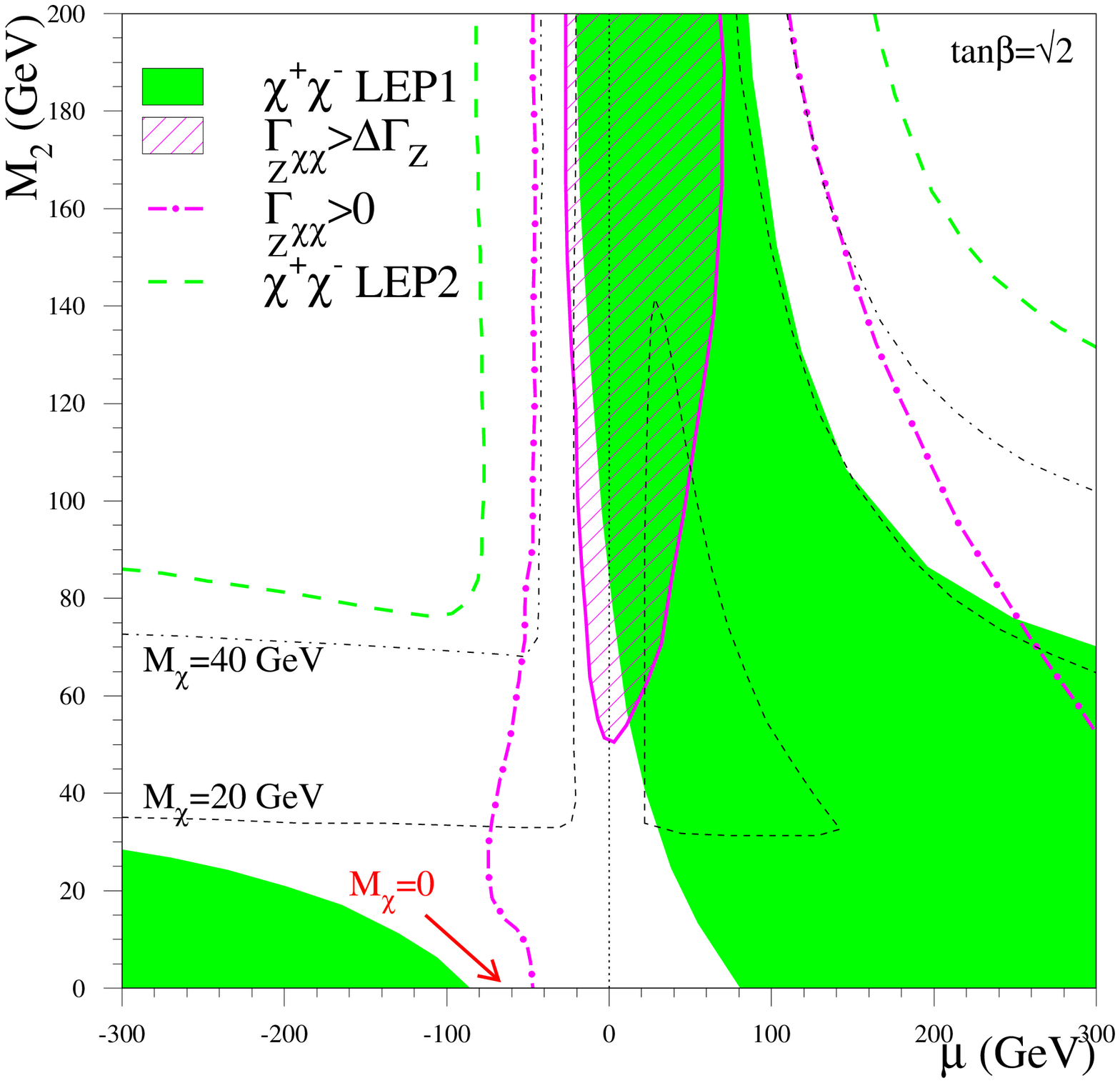}} 
  \end{tabular}
  \caption{\em {\small  
  Left:Portions of the parameter space $(\mu,\Mtwo)$ giving the relic 
  density in the cosmologically preferred range 0.1$<\mathrm{\Omega_{CDM}h^2}<$0.3 
  (light gray).
  Higgs boson and chargino mass contours are shown as dot-dashed and dashed lines, 
  respectively. Here $\tanb=3$, $\mzero=1000$ $\Gcsq$, $\mA=1000$ $\Gcsq$. 
  From Ellis 2000a).
  Right: Basic features of the $(\mu,\Mtwo)$ plane for $\tanb=\sqrt{2}$.}}
 \end{center}
\end{figure}

\section{LEP and its detectors}

The LEP machine is a synchrotron electron-positron collider located at CERN 
in the surroundings of Geneva, Switzerland. 
The first operation phase (LEP~1, 1989-1995, $\rts=88\div95$~\gev, 
$\sim$175 pb$^{-1}$/experiment collected) was dedicated to the precise 
measurement of the Z resonance. During the second phase, just completed, 
(LEP~2, 1995-2000, $\rts=130\div209$~\gev, $\sim$650 
pb$^{-1}$/experiment collected) 
the machine was optimized for new particle discovery. 
The results discussed here are based on the data collected before 2000 
at $\rts\leq 202$~$\gev$. 

The four detectors installed in the interaction points were designed 
for precision measurements. They feature a large covering of the solid 
angle (typically 80\% for charged particles, 95\% for neutrals), good 
lepton (electron, muon) identification, photon reconstruction and 
heavy flavor (b,$\tau^\pm$) tagging, good hadron jet reconstruction 
capabilities. The total visible energy $\evis$ is usually measured with a
resolution of 10\%$\cdot\sqrt{\evis}$. Finally, the trigger efficiency 
is typically 100\% provided $\evis>\sim 3$ GeV. 

During the LEP~1 phase, a precise scan of the Z resonance was performed, 
allowing  a precise determination of the Z parameters to be done. 
The parameters of the Z boson, in particular the Z widths, 
are expected to be modified 
by the opening of the new potential decay channels predicted in the MSSM. 
By comparing the most recent determination of the Z widths (LEP EW 2000) with the 
predictions, 95\% confidence level upper limits of 6.2 and 1.7 MeV on new 
contributions to the total and invisible 
Z widths, respectively, can be derived. 
These exclude Heavy Dirac or Majorana neutrinos and sneutrinos masses  
up to about 40 GeV/c$^2$. In particular, the precision measurements
 definitely rule out the sneutrino 
as supersymmetric candidate for CDM (see Sect.~2), representing a good example 
of complementarity between indirect and direct searches.

\section{Constraining the neutralino: basic idea}

The precise measurements of the Z resonance do not set absolute 
constraints on the parameters of the $\PChiz{1}$ because 
it is possible to find configurations 
in the parameter space where the coupling associated with Z$\PChiz{1}\PChiz{1}$ vertex 
is very small. Direct search for $\PChiz{1}\PChiz{1}$ events would of course not 
help since they would not be recorded by the detectors. 
In such cases, sensitivity to $\PChiz{1}$ parameters can be recovered by looking at other 
processes predicted by the model for a given set of parameters.
In this section we present the basic ideas underlying the analysis; 
more details are given in Sect.~5.
The GUT relations introduced in Section~2 (Eq.~1) establish 
a link between $\Mone$ and $\Mtwo$, 
$\Mone = \frac{\alpha_1(\mZ)}{\alpha_2(\mZ)} 
\Mtwo \sim \frac{\Mtwo}{2}$, 
so that the properties of both the chargino and neutralino 
sectors are described in terms of $\tanb$, $\mu$ and $\Mtwo$. 
Exact formulae for mass matrices can be found in the literature 
(see, for instance, Haber 1985). 
Of relevance here is the approximate relation 
$\Mchi \propto \mathrm{Min}(\Mtwo,|\mu|)$, which implies 
that an absolute sensitivity to $\Mchi$ requires sensitivity 
to the model predictions for $\Mtwo,|\mu|\!\sim\!0$
\footnote{Though both $\lsp$ mass and couplings are relevant for 
the calculation of $\ocdm h^2$, 
the CDM constraints are usually given in terms of an absolute lower 
limit on $\Mlsp$, valid, under the given assumptions, for all the field configurations. }.
As it was first noticed in Ellis 1996, the common gaugino description 
can be used to derive constraints on the neutralino sector from the results of 
searches for chargino pair-production.
This works particularly well when the sfermions are heavy, i.e. out of 
the LEP kinematic reach. In such a case (see Sect.~5.1) 
the LEP data exclude chargino pair production essentially 
up to the kinematic limit; at $\rts\ge 130$ ~$\gev$ this implies 
that parameter settings with $\mu\!=\!0$ or 
$\Mtwo\!=\!0$ are incompatible with the data (see Fig.~1b), 
thus allowing an absolute lower limit on $\Mlsp$ to be derived    
\footnote{From Fig.~1b we see, {\it en passant}, that the kinematic reach of LEP~1 
was not large 
enough to allow setting an absolute limit on $\Mlsp$ at low $\tanb$ (Ellis 1996)}.  
In the case the sfermions are light, the reach 
of chargino searches is reduced (Sect.~5.2). However, the sfermions, in particularly the 
sleptons, could be light enough to be in the LEP kinematic reach. 
The constraints of sleptons pair-production can be translated into 
constraints on the gaugino sector using 
the GUT relations for sfermion masses (Eq.~2); 
the loss of sensitivity of chargino searches is then partially backed up (Sect.~5.3). 
Finally, pushing further the use of the GUT relations for sfermion masses, 
stronger constraints on $\Mlsp$ can be derived using 
the limits inferred on the squark sector from the 
negative results of Higgs boson searches (Sect.~5.4).

\section{Direct searches at LEP 2}

The analysis discussed in Sect.~3 is an example of 
indirect search for new phenomena. The {\it standard} processes, 
which constitute the {\it background} for the new physics ones, are 
measured as precisely as possible, and the result is compared 
with the prediction of the SM, looking for deviations. 
The sensitivity $s$ to deviations is well approximated by $s= S/\sqrt{B}$ , 
being $S$ the signal yield expected and the $B$ the residual 
background, and is therefore limited by the statistical fluctuations 
of the background which are typically large, 
since no attempt is done in the indirect searches to reduce it. 
The purpose of direct searches is to try to increase the sensitivity 
of the search by reducing the background contamination keeping 
reasonable efficiency on the signal, so that $s$ is increased. 
For practical reasons, 
most of the results used in this section are taken from the ALEPH 
publications (ALEPH 1998 and ALEPH 2000); however similar results 
have been obtained by the other LEP collaborations 
(DELPHI 2000b, L3 2000, OPAL 2000). 

Production in $\epem$ collisions proceeds via $s$-channel exchange 
of a Z boson and, when allowed by quantum numbers, $t$-channel exchange of a 
supersymmetric particle (Fig.~2a,b). The relevance of the latter can be judged 
from Table~2, which gives the production cross-sections close 
to threshold and the $\rts$ dependence 
for a few representative processes. Chargino production has the largest
cross-section, whatever the field composition, up to masses very close to 
the kinematical limit. However, the effect of a light sneutrino 
$t$-channel exchange, interfering destructively with the $s$-channel exchange, 
is significant, in particular in the so called gaugino region ($\Pcha\sim\tilde{\mathrm{W}}^\pm$). 
For neutralinos and sleptons the interference can be 
constructive. 

\begin{table}[thb]
\label{Tab_xsec}
 \begin{center}
\renewcommand{\arraystretch}{1.}
\begin{tabular}{|c|c|c|c|} 
\hline
Process & f($\bkin$)   & at $\sqrt{s}=200$ GeV & Comments  \\
\hline \hline
$\Pccp\Pccm$ & $\bkin$  &  2.1 pb         &  $\Pcha\sim\tilde{\mathrm{W}}^\pm$, heavy sneutrino \\
             &          &  1.1 pb         &  $\Pcha\sim\tilde{\mathrm{H}}^\pm$, heavy sneutrino \\
             &          &  0.5 pb         &  $\Pcha\sim\tilde{\mathrm{W}}^\pm$, $\Msnu=50$~\Gcsq \\
             &          &  0.9 pb         &  $\Pcha\sim\tilde{\mathrm{H}}^\pm$, $\Msnu=50$~\Gcsq  \\
\hline 
$\PChiz{2}\PChiz{1}$ & $\bkin$  & 0.4 pb    & $\tanb\!=\!4,\mu\!=\!-95\ \Gcsq,\Mtwo\!=\!500\ \Gcsq$ \\
           & &   & $m_{\PChiz{2}}\!+\!m_{\PChiz{1}}=195\ \Gcsq$, heavy selectron \\
\hline
$\PSe^+_{\mathrm{R}}\PSe^-_{\mathrm{R}}$ & $\bkincube$ &   0.13 pb &  
      $\tanb\!=\!2,\mu\!=\!-100\ \Gcsq,\Mtwo\!=\!100\ \Gcsq$ \\
\hline
$\PSmu^+_{\mathrm{R}}\PSmu^-_{\mathrm{R}}$ & $\bkincube$ &   0.04 pb &  
       \\
\hline 
\end{tabular}
  \caption{\em {\small Example cross sections. The produced particle mass is 
 95 $\Gcsq$ unless otherwise indicated.} }
 \end{center} 
\renewcommand{\arraystretch}{1.}
\end{table}

\begin{figure}[thb]
\label{Fig_susysig}
 \begin{center}
  \mbox{\epsfig{file=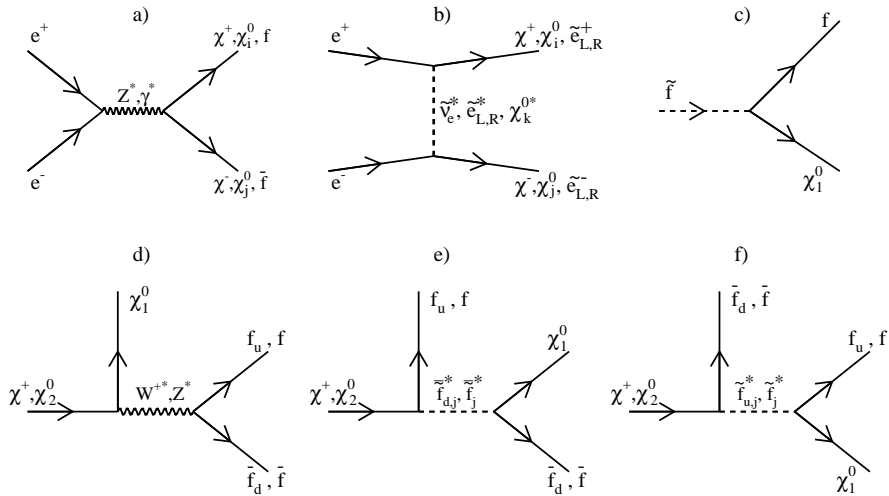,width=12cm}}
  \caption{\em {\small Lowest order graphs for the relevant signal processes.}}
 \end{center}
\end{figure}

The decay patterns are depicted in Figs.~\ref{Fig_susysig}d-f.
 For heavy sfermions, the decays 
of charginos and neutralinos are dominated by the exchange of  
W and Z, respectively, which at LEP energies are off-shell (3-body decays).
 The final states and the branching fractions will 
reflect those of the heavy bosonic mediators, thus with a predominance of hadronic 
final states. The graphs involving sfermions become of increasing 
importance as the sfermion masses decrease; eventually, when a sfermion is 
light enough to be produced on-shell in the decay (2-body decays) the 
corresponding process largely dominates over the 3-body decay processes.

We turn now back to the phenomenology of the final states.
The $\lsp$'s interact only weakly and therefore escape detection. 
The fact that at least two of them are present in the final state 
gives apparent energy non-conservation, i.e. {\it missing energy}.
This is the well known signature for supersymmetry under R-parity 
conservation.   
The {\it visible energy}, i.e. the energy carried by particles that 
can be detected, is directly related to the difference 
in mass $\dm\!=\!M\!-\!\Mlsp$ between the produced particle and the $\lsp$.  
Under the assumption of two body decays where the visible system 
has negligible mass, the visible energy can be approximated by 
\be
\evis \simeq \frac{\sqrt{s}\dm}{4 M} \left( 1+\frac{\Mlsp}{M} \right)
\ee
There is, therefore, an intrinsic problem in detecting the signal when
$\dm\to 0$, partially recovered by the boost for very small $\dm$ 
(like in $\toto$ production, for instance), thanks to the Lorentz boost. 
Since the trigger needs at least 3-5 GeV of visible energy to have 
reasonable efficiency, we see from Eq.~3 that for particles 
close to the kinematic limit, direct searches will be blind 
for $\dm\leq\!2\!-\!3$ GeV, unless very special trigger techniques 
are adopted. 
However, for the CDM constraints discussed in this paper, the problem 
is less severe, since, in the interesting regions (see Fig.~1b),  
the $\lsp$ is gaugino like with typically large $\dm$'s.  

\subsection{Backgrounds}

The background processes are those processes predicted by the standard 
model which can fake the signal, i.e. can have missing energy. In order 
of decreasing cross section these are:

\boldmath$\ggff$\unboldmath. 
         The so-called {\it gamma-gamma} processes described at 
         tree level by the graph shown in Fig.~3a. The initial state 
         electron and positron are usually scattered at very low angle 
         remaining undetected in the beam pipe. The remaining system can 
         feature energy-momentum imbalance; however, its energy is typically 
         small and the corresponding momentum non conservation limited. 
         These processes have huge cross section ($\sim$nb) but they 
         can be effectively reduced, and represent a problem only at very 
         low $\dm$. 

\boldmath$\epem\!\to\!\ffbar(\gamma)$\unboldmath. 
         These are shown in Fig.~3b and can 
         have large cross sections (up to 100 pb). Except for the case 
         f=$\nu$, which is not relevant here, these events  can 
         fake momentum imbalance only one final state particles is not well measured. 
         In such a case they are reduced by requiring the direction of the missing 
         momentum to be away from passive regions of the detector, like the 
         beam line, for instance. 

\boldmath$\ffff$\unboldmath. 
         These are mediated by heavy gauge boson exchange and are
         dominated by $\epem\!\to\!\ww$ (Fig.~3c,d). They have signal like 
         cross sections (up to $\sim$20 pb), and when $\mathrm{f_i}=\nu$ 
         they can feature very large transverse momentum (with respect to the beam axis)
         and missing energy, resulting 
         very signal-like at large $\dm$. Usually they can not be eliminated 
         but have to be treated statistically in the evaluation of the 
         result. 

\begin{figure}[thb]
\label{Fig_bkgproc}
 \begin{center}
  \mbox{\epsfig{file=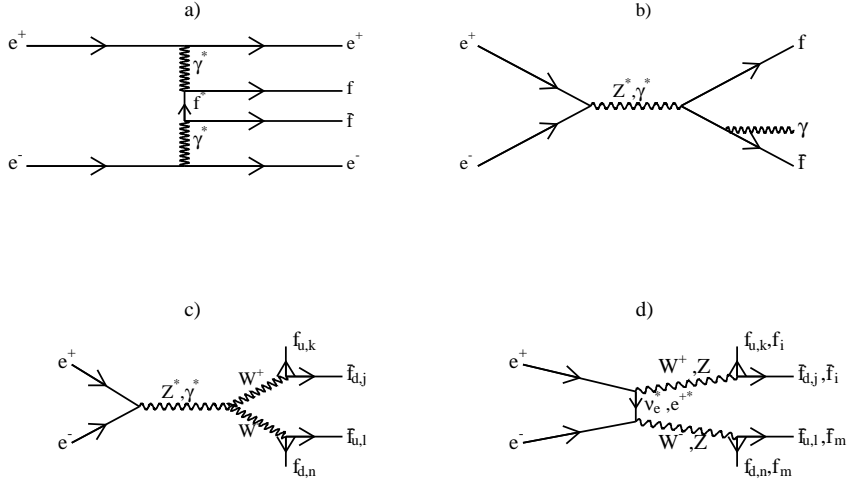,width=12cm}}
  \caption{\em {\small Graphs of the main background processes.}}
 \end{center}
\end{figure}

\subsection{The main process: $\epem\to\chi^+\chi^-$ with heavy sfermions}
We start the discussion 
by assuming that the sfermions are heavy enough to make their effect on 
the cross section and decay processes negligible. For charginos in the reach 
of LEP~2 energies this corresponds to $\mzero\ge 200$~GeV if the GUT 
relations Eq.~2 are used for the sfermion masses. 
As we have seen above, 
the main decay process in this case is $\mathrm{\Pcha\!\to\! (W^{\pm})^*\PChiz{1}}$ 
giving topologies similar to $\ww$, with additional missing energy. 
The three main topologies searched for are given in Table~3 together with 
the main variables used to discriminate the signal, the residual background 
and the typical efficiency. 

\begin{table}[thb]
\label{Tab_chargino_res}
 \begin{center}
\renewcommand{\arraystretch}{1.1}
\begin{tabular}{|c|c|c|c|c|c|} 
\hline
Signal & Topology      & Discriminating & $\epsilon$  & \multicolumn{2}{|c|}{Background} \\
\cline{5-6}
$\Pccp\Pccm\!\to$    &          & variables & (\%) &  processes & $\sigma_{bkg}$ \\
\hline \hline
$\lsp\lsp+\Pu_i\Pad_j\Pau_n\Pd_m$ & 
                       jets + $\notE$ & $\PT$,$\mvis$ &  $\sim$35  & $\znn$, $\wen$, &  $\sim$45 fb   \\
   &                     &               &            & $\ggqq$   &             \\
\hline
$\lsp\lsp+\Pu_i\Pad_j\bar\nu\ell^-$   & jets+$\ell^\pm+\notE$ & $\PT$,$\mvis$,$\ell^\pm$ energy, &  $\sim$50  & $\ww$, $\zz$, &  $\sim$25 fb   \\
    + {\it h.c.}      &            & $\ell^\pm$ isolation & & $\ggqq$    &             \\
\hline
$\lsp\lsp+\nu\ell^+\bar\nu\ell^-$   & $\ell^+\ell^-+\notE$  & $\PT$,$\mvis$, &  $\sim$45  & $\ww$, $\znn$,  &  $\sim$150 fb \\
          &            & acoplanarity  &            & $\gagall$   &             \\
\hline 
\end{tabular}
  \caption{\em {\small Typical performance of chargino searches at large $\mzero$.
Here $\PT$ is the total momentum transverse to the beam, $\mvis$ the visible 
mass, $\ell^\pm$ an identified electron or muon; ``jet'' indicates collimated 
clusters of particles 
originating from the hadronization of quarks. } }
 \end{center} 
\renewcommand{\arraystretch}{1.}
\end{table} 

The absence of any excess above the standard mode expectation in the data 
collected up to 202 GeV (ALEPH 2000, DELPHI 2000b, L3 2000, OPAL 2000), 
allows each collaboration to 
set upper limits on the cross section for the specific process of the order 
of $(0.1\div 1)$ pb, depending on the chargino mass and on $\dm$. 
Translating in the ($\mu,\Mtwo$) plane for fixed $\tanb$ this result is sufficient 
to exclude chargino pair production up to the kinematic limit. This is shown, 
for example, in 
Fig.~4a for $\tanb=\sqrt{2}$.  In particular the axis of the plane are excluded, 
which means that a lower limit on the $\lsp$ mass can be derived for each 
$\tanb$ value as shown in Fig.~4b. 
The result is basically determined by the kinematic limit for chargino 
production, and therefore is limited by the maximum centre-of-mass energy reached. 
It turns out that the lowest $\lsp$ mass is always reached in the gaugino-mixed 
regions of the ($\mu,\Mtwo$) plane, where the relation $\Mchi\sim\Mcha/2$ is 
approximatively valid. The increment with $\sqrt{s}$ of the mass limit is therefore 
given by $\Delta\sqrt{s}/4$, as can be verified in Fig.~4b. 

\begin{figure}[thb]
\label{Fig_mum2small}
 \begin{center}
 \begin{tabular}{cc}
  \mbox{\epsfysize=8cm \epsffile{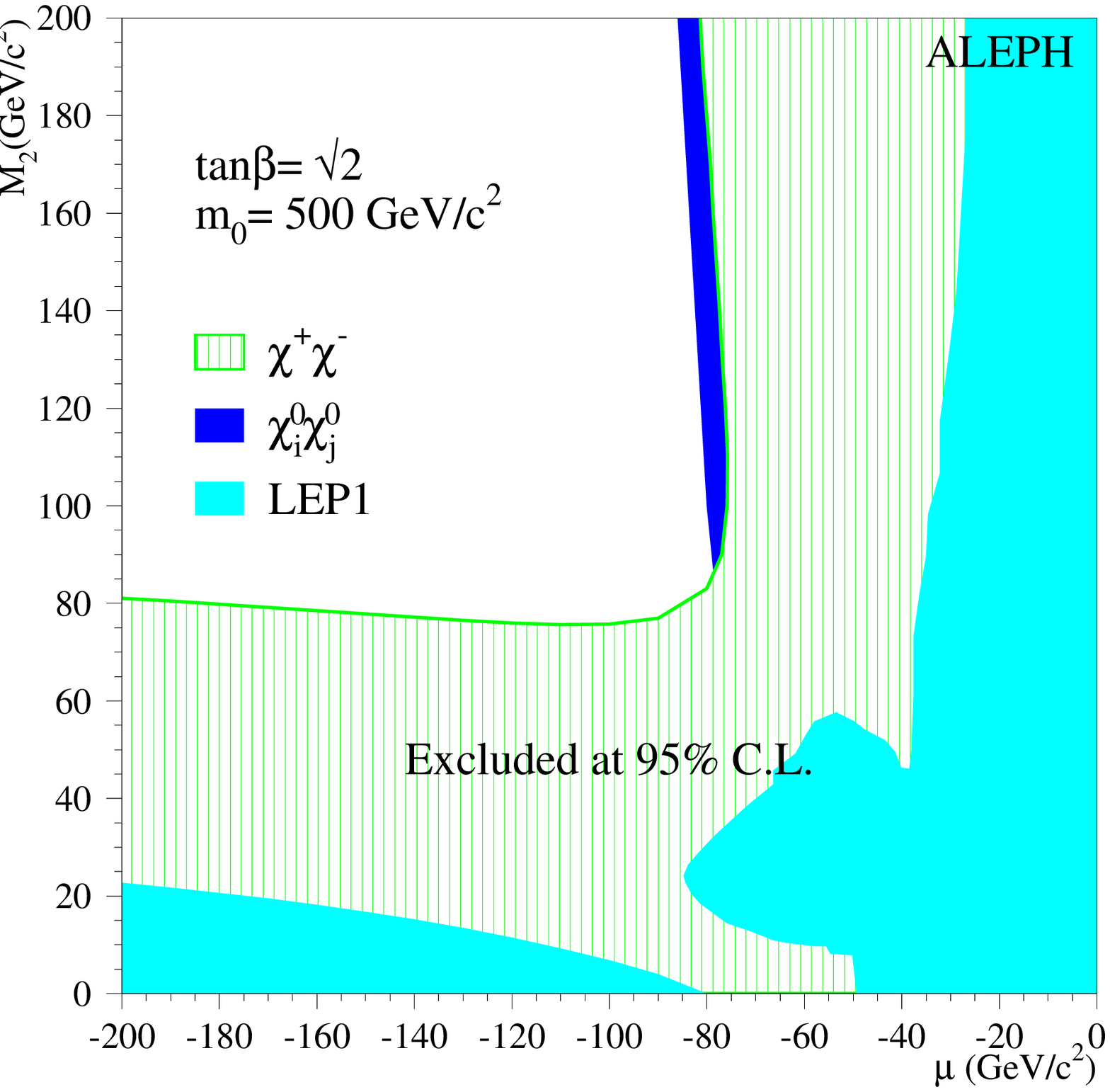}} &
  \mbox{\epsfysize=8cm \epsffile{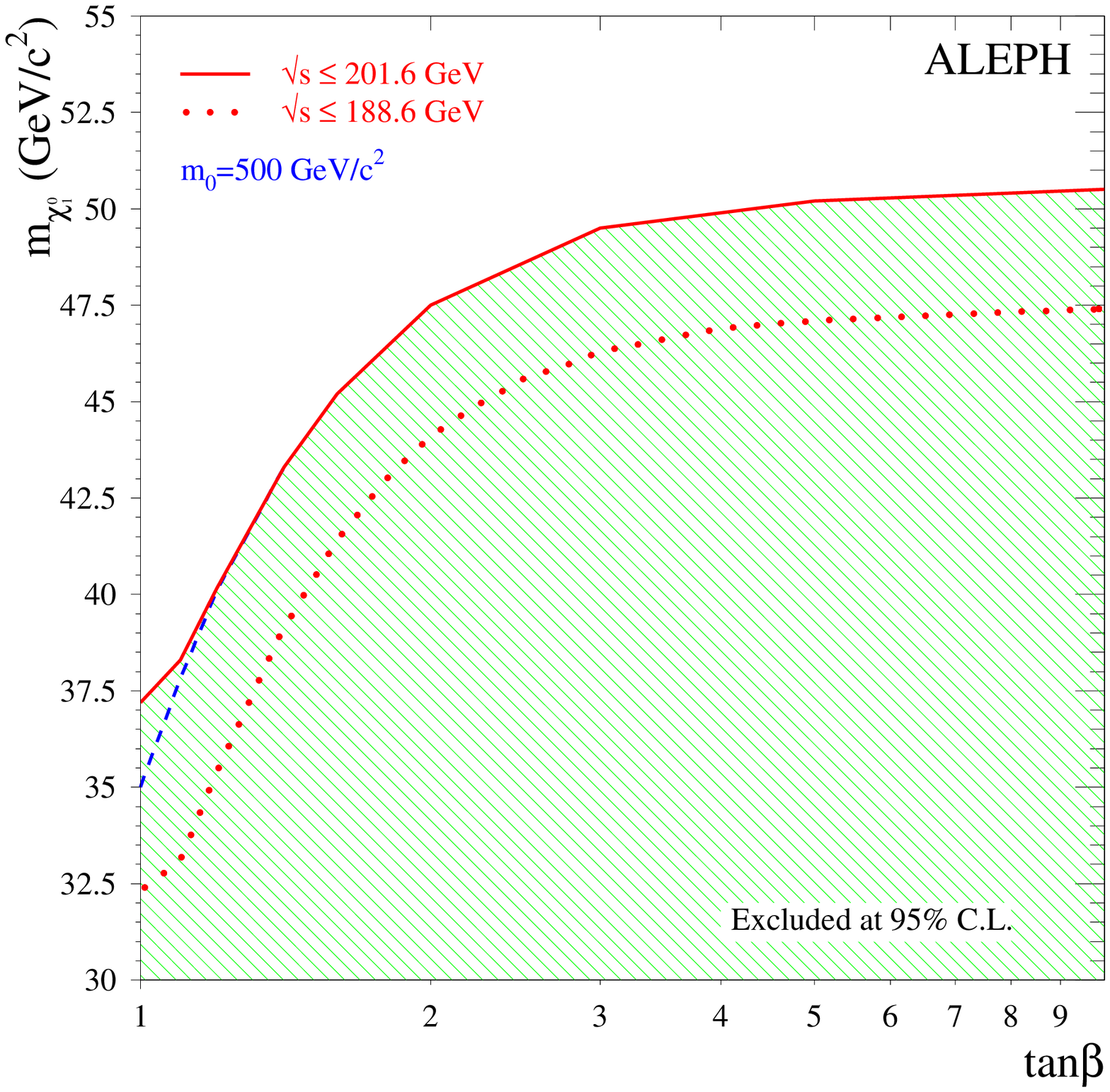}}\\
  \mbox{\epsfysize=8cm \epsffile{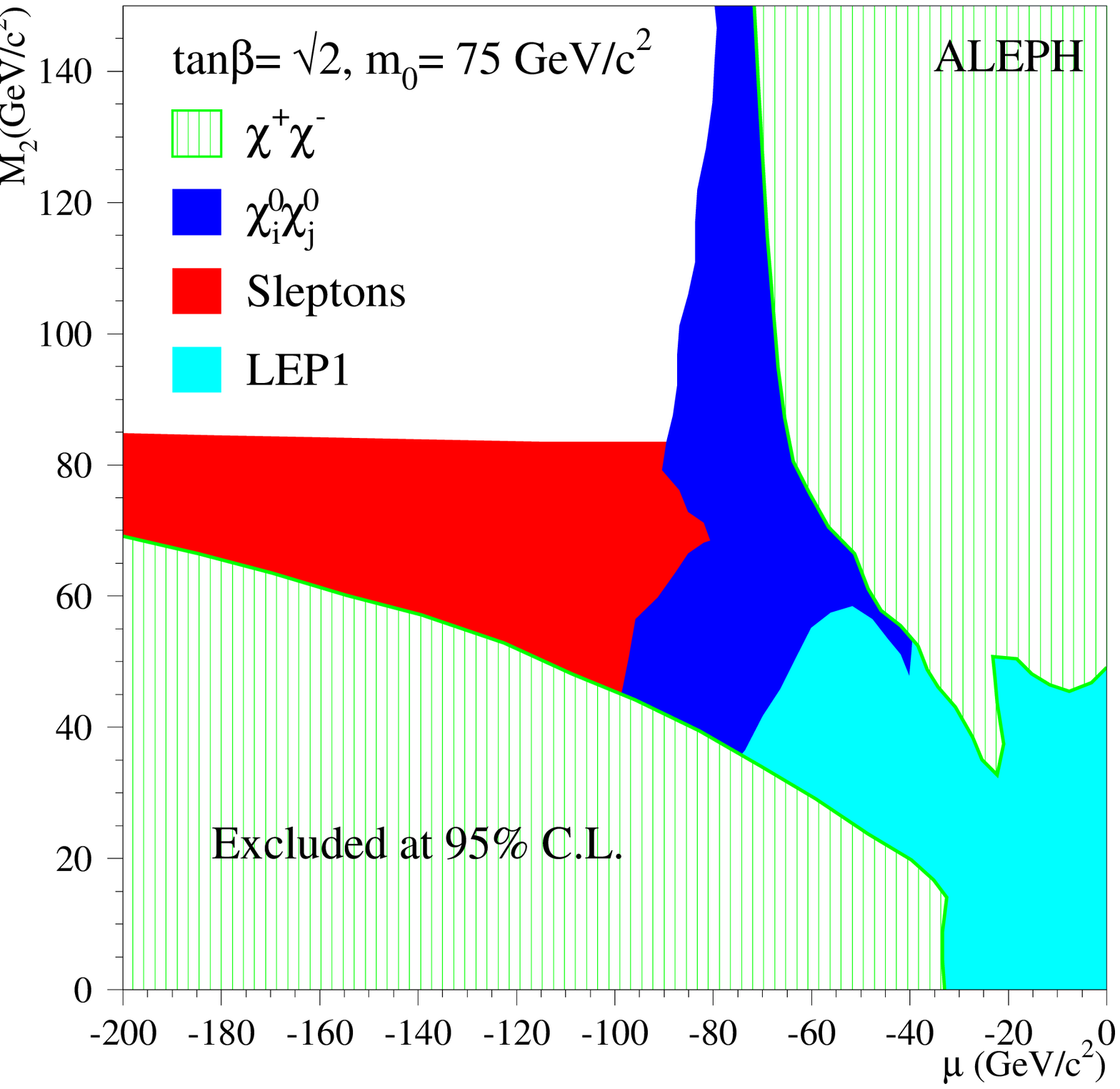}} &
  \mbox{\epsfysize=8cm \epsffile{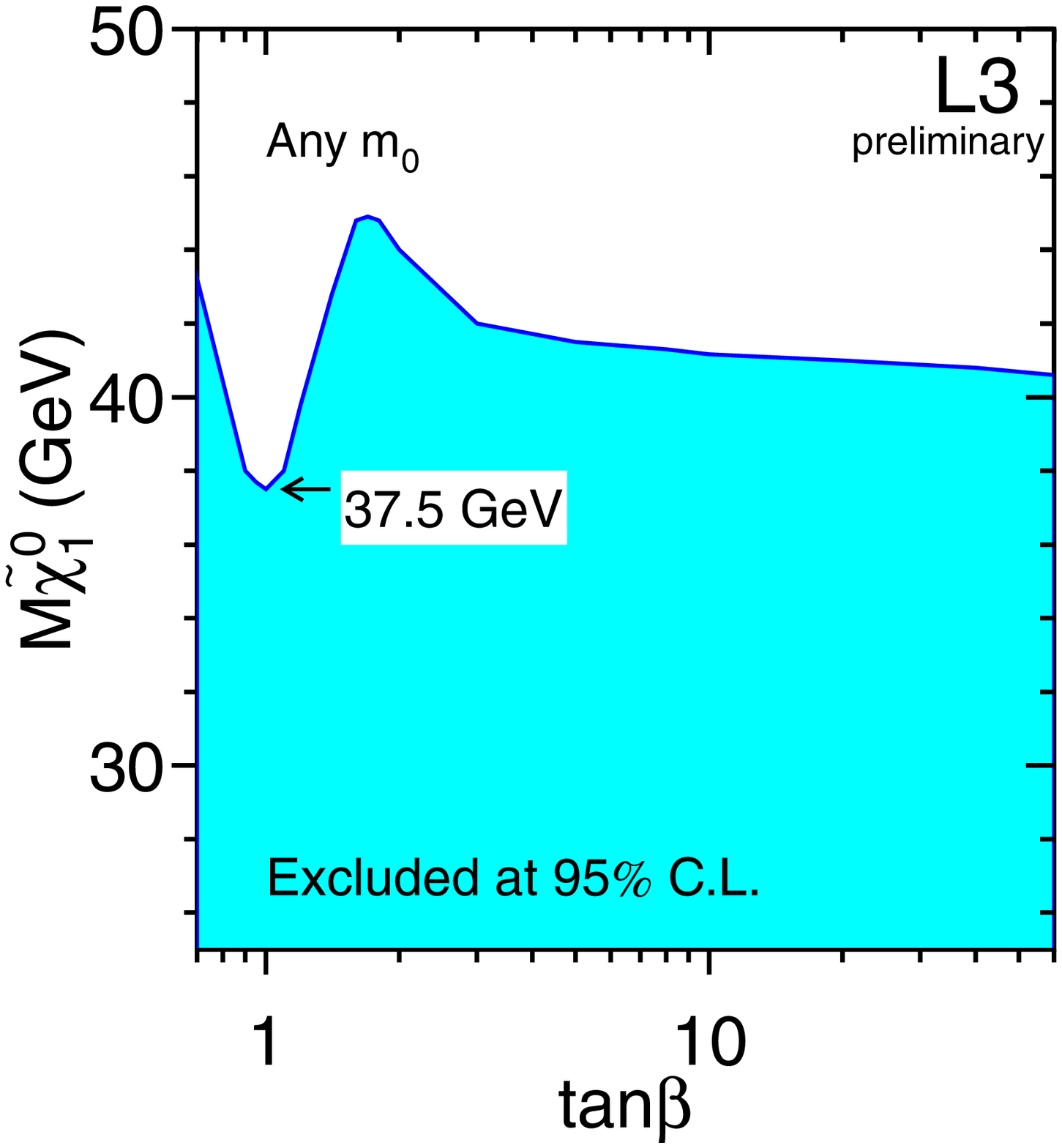}} \\
 \end{tabular}
  \caption{\em {\small Top-Left: Regions excluded by ALEPH in the mixed region 
   of the ($\mu,\Mtwo$) plane  
  for $\tanb=\sqrt{2}$ using data collected up to $\rts$=202 GeV and 
  under the assumption of heavy sfermions (Ref. ALEPH 2000). 
    Top-Right: The lower limit on $\Mlsp$ as a function of $\tanb$ set by ALEPH 
  under the assumption of heavy sfermions using data collected up to $\rts$=202 GeV. 
  The result obtained using the data up to $\rts=189$ GeV are shown for 
  comparison (Ref. ALEPH 2000).
  Bottom-Left: Light sfermions scenario: sensitivity in the ($\mu,M_2$) plane  
 of the ALEPH searches at  $\rts\leq 189$~\gev for $\mzero=75$~\Gcsq  
  (Ref. ALEPH 1998). 
  Bottom-Right: The lower limit on $\Mlsp$ as a function of $\tanb$ set by L3  
  valid for any value of the common sfermion mass $\mzero$ obtained using data
  collected up to $\rts$=202 GeV (Ref. L3 2000).} }
 \end{center}
\end{figure}

The neutralino processes give final states topologically not much different 
from those predicted for charginos, which are selected with similar 
performances. However, since the cross sections are much smaller, their impact 
in the combined interpretation is visible only in a corner of the parameter 
space, namely at low $\tanb$ in the mixed region (Fig.~4b).

\subsection{The light sfermion case}
The GUT relations (Eq.~2) predict the sleptons lighter than the 
squarks and therefore to have the larger effect of gaugino phenomenology 
at LEP. 
Mixing in the stau sector could lead to some weird effect (LEP SUSY WG 1999); 
there is no systematic study of this available from the LEP collaborations. 
However there is enough information (DELPHI 2000a) to believe that even 
those bizarre complications 
could be dealt with without invalidating the general 
picture discussed here. 

The qualitative effect of decreasing the sfermion masses is displayed in 
Fig.~4c, showing  the ALEPH result for $\rts\leq189$~\gev\ 
and $\mzero=75\ \Gcsq$. A qualitative comparison with Fig.~4a 
indicates that: {\it i)} the chargino sensitivity is reduced especially in the 
mixed region because of the smaller cross-section and the opening of 2-body 
decay channels that for very low $\dm$ lead to invisible final states; 
{\it ii)} the neutralino 
sensitivity somewhat enhanced for relatively large $\Mtwo$ because of the 
larger cross-section, but it strongly decreased for large negative $\mu$ and 
relatively small $\Mtwo$ because of the invisibility of the decay $\PChiz{j}\to\PSnu\nu$ 
which dominates in that region for low $\mzero$. The net result is that in the 
mixed region the sensitivity of the combined gaugino searches is reduced. 

The inclusion of the sleptons using the GUT relations allows to recover 
the sensitivity lost by gaugino searches. 
Sleptons have been searched for mainly through the standard decay channel 
$\PSl\to\lsp\ell$, producing pairs of acoplanar leptons, and  
they would be selected with rather high efficiencies and relatively low 
backgrounds by all the four LEP detectors, in particular for the selectron 
and smuon channels. 
Cross section upper limits of the order of 0.025 pb, 
usually shown in the plane ($\Msl$,$\Mlsp$), have been 
obtained by the combination of the results. 
From these upper limits, mass lower limits are derived, which can be superimposed 
to the gaugino constraints. For example, the ALEPH lower limit for 
$\Mlsp\simeq40$~$\Gcsq$ is about 93~$\Gcsq$ (ALEPH 2000). 
Using the GUT relations (Eq.~2) and assuming 
no mixing in the stau sector, results like the one shown in Fig.~4c are 
obtained. 
The study of plots like these 
for different $\tanb$ and $\mzero$ values 
allows the derivation of a lower limit of $\Mlsp$ independent of $\mzero$. 
The L3 result is shown as a function of $\tanb$ in Fig.~4d (L3 2000). 

Two comments about the figure. For $\tanb$ close to 1 the sensitivity obtained 
at small $\mzero$ including the sleptons is better than the one reached at 
large $\mzero$ by gaugino searches alone. In that $\tanb$ region Fig.~4d 
is therefore the same as Fig.~4a. 
For $\tanb>\sim 2$ the sleptons are not able to cover completely the {\it corridor}, 
the low $\dm(\!=\!\Mcha\!-\!\Msnu)$ region where charginos are invisible. 
The limit shown in such case 
is obtained at small $\mzero$ by slepton searches. However, this limit is better 
than the one obtained at large $\mzero$ for $\tanb=1$. 

The absolute lower limit on $\Mlsp$ using data collected at centre-of-mass energies 
up to 202 $\gev$ is 37.5 $\Gcsq$, reached at large $\mzero$ for $\tanb=1$.
This result has been obtained using the lowest order GUT relations for 
$r=\Mone/\Mtwo$ at electroweak scale, calculated assuming no 
physics between the supersymmetry breaking and the grand unified scales, and 
no radiative corrections to gaugino masses. 
Even assuming that no new physics other than supersymmetry appears below the GUT 
scale, one should consider the inclusion of high order corrections to $r$ 
and radiative corrections to the gaugino masses, each expected to affect by about 
3\% the lower limit on $\Mlsp$. 

\section{Impact of Higgs searches}

Finally we would like to comment on the impact of the searches for Higgs bosons. 
Under the assumption of GUT relations for sfermion masses, a link between 
the Higgs and gaugino sectors (absent at tree level) is introduced via the 
those squarks which are partners of heavy quarks, like the stops and sbottoms. 
These in fact induce large radiative corrections to the Higgs sector 
(see, for instance, Ellis 1991). 
The mass of the lighter 
state $h$ is predicted to be $\leq 150$~$\Gcsq$ and extensive searches have 
been performed at LEP for the lighter Higgs boson states, in particular in the 
channels $\epem\to h Z$ and $\epem\to h A$. From the negative results of such 
searches a lower limit of the order of 110~$\Gcsq$ 
has been set of the mass of the lighter neutral scalar Higgs boson
\footnote{During 2000 a hint for a 115 $\Gcsq$ mass Higgs boson has been 
found by the LEP experiments (LEPC 2000); the possible implications 
of such findings on supersymmetric CDM are discussed, for instance, in Ellis 2000b.}. 
Setting  a lower limit on $\mh$ translates into a lower limit, for example,
on $\mst$, which in turn can be interpreted as a constraint in the 
($\mzero,M_2$) plane via the relevant Eq.~2, and combined 
with the constraints from gaugino searches as in the case of 
sleptons.
Of course the result will depend not only on $\tanb$ but also on the chosen 
value for $\mA$ and $A_\PSt$. 
A scan over these two parameters has shown that
the strength of the Higgs results are such 
that at low $\tanb$ there is indeed an improvement in the sensitivity to
$\Mlsp$, acting as a cut-off at $\tanb\simeq 2$ (see ALEPH 2000). 
However, there is no general improvement at large $\tanb$ were the limit is 
set in the corridor by the sleptons searches. 

\section{Conclusions}

Under the assumptions leading to the simplest GUT relations for 
the supersymmetric parameters and of R-parity conservation, the LEP 
searches for supersymmetry allowed to set an absolute lower limit 
of 37.5 $\gev$ on 
the mass of the best accredited supersymmetric candidate for 
CDM, the lightest neutralino $\lsp$, 
using the data collected up to 202 $\gev$. These kind of 
analysies complement the direct searches for CDM, usually 
less sensitive or even insensitive to very low WIMP masses. 
The absolute lower limit suffers of theoretical uncertainties of the 
order of a few $\gev$, and is mainly determined by the kinematic 
limit for chargino pair production (at small $\tanb$) and the constraints 
on slepton pair production (at large $\tanb$). 
Stronger constraints can be obtained at small $\tanb$ by the inclusions 
of the results of Higgs bosons searches.
Finally, the interpretation of  
the non-observation of signals for supersymmetry 
in the last sample of LEP data collected during the year 2000 (LEPC 2000), 
is not expected to improve significantly on the present results; hence  
the results persented here can be considered to be a good approximation 
of the final ones.

\paragraph {Acknowledgements.}
I am grateful to M.~Maggi and S.~Savaglio for the careful 
reading of the manuscript, and to the organizers for inviting me  
to the interesting workshop.

\section { References}

\reff ALEPH Collaboration: 1998, 
Euro. Phys. J. {\bf C 11} (1999) 193.

\reff ALEPH Collaboration: 2000, 
         CERN EP/2000-139, sub. to Phys. Lett. B.

\reff Bottino, A. and Fornengo, N.: 1999,  
       hep-ph/9904469. 

\reff Caldwell, D.O. et al.: 1988, 
      Phys. Rev. Lett. {\bf 61} (1988) 510. 

\reff DELPHI Collaboration: 2000a, 
       CERN EP/2000-133

\reff DELPHI Collaboration: 2000b, 
       DELPHI 2000-072, sub. to XXX ICHEP, Osaka, 2000.

\reff Ellis, J. et al.: 1983, 
       Nucl. Phys. {\bf B 238} (1984) 453. 

\reff Ellis, J., Ridolfi, G., Zwirner, F.: 1991, 
      Phys. Lett. {\bf B 257} (1991) 83.  

\reff Ellis, J., Falk, T., Olive, K., Schmitt, M.: 1996, 
                 Phys. Lett. {\bf B 388} (1996) 97.

\reff Ellis, J.: 1998, 
                 Phys. Scripta {\bf T 85} (2000) 221, astro-ph/9812211. 

\reff Ellis, J., Falk, T., Ganis, G., Olive, K.: 2000a, 
                 Phys. Rev. {\bf D 62} (2000) 075010. 

\reff Ellis,J.,~Ganis,G.,~Nanopoulos,D.V.,~Olive,K.: 2000b, 
          hep-ph/0009355, sub.~to~Phys.Lett.B.

\reff Feng, J.L., Matchev, K.T., Wilczek, F.: 2000,
             astro-ph/0008115, sub. to Phys.Rev.D. 

\reff Haber, H., Kane, G.: 1985, 
              Phys. Rept. {\bf 117} (1985) 117 

\reff L3 Collaboration: 2000, L3 note 2583, sub. to XXX ICHEP, Osaka, 2000. 

\reff LEPC: 2000, Public Minutes of Sessions held on Sept. 5, Oct. 10, Nov. 3, 2000. 

\reff LEP EW WG: 2000, 
           CERN EP/2000-153 

\reff LEP SUSY WG: 1999, 
          note LEPSUSYWG/99-03.1, 
     {\tt{\small http://lepsusy.web.cern.ch/lepsusy/}}

\reff Martin, S.: 1997, 
                 in {\it Perspectives on Supersymmetry}, Ed. G.~Kane,
          hep-ph/9709356. 

\reff Nilles, H.P.: 1984, 
              Phys. Rept. {\bf 110} (1984) 1. 

\reff OPAL Collaboration: 2000, 
       OPAL PN 418, submitted to XXX ICHEP, Osaka, 2000.

\reff Turner, M. : 1999, in proceedings of PIC99, Ann Arbor, astro-ph/9912211. 

\end{document}